\documentclass[a4paper]{article}
\begin{document}
\newcommand{\beq}{\begin{equation}}
\newcommand{\eeq}{\end{equation}}
\newcommand{\gfrc}[2]{\mbox{$ {\textstyle{\frac{#1}{#2} }\displaystyle}$}} 
\title{Mass formulae for spherically symmetric stellar configurations in five diemsional space time}
\author{ G S Khadekar\footnote{e-mail: gkhadekar@rediffmail.com}, K S Adhao\footnote{e-mail:ati-ksadhav@sauharnet.in} and H R Ghate }

{ {\renewcommand{\thefootnote}{\fnsymbol{footnote}}

\footnotetext{\kern-15.3pt AMS Subject Classification: 83D05, 83 Exx, 83F05 }}
\maketitle

\begin{abstract}
 An expression is derived where the mass is connected to an integral over the pressure of gravitating matter in the frame work of five dimensional (5D) space time. \par
\end{abstract}

\section { Introduction} 
In view of recent developments in superstring theroy and ten dimensional $N=1$ Yang-Mills supergravity in its field theory limit, need higher dimensional space time is incerasing. For this reason, in the recent years there has been considerable interest in theories with higher dimensinal space-times in which extra dimesions are contracted to a very small size, apparently beyond our ability for measurment. Marciano (1984) has pointed out that the experimental detections of time variation of fundamental constants should be strong evidence for the existence of extra dimensions. Since the world around us is manifestly 4D in the present era, both in supersymmetry and Kaluza-Klein ( K K )  theories we need to understand the mechanisms of reduction in size of the extra dimensions in order to realise the 4D world. Kaluza-Klein theories are interesting approaches to unify gravity and guage fields, while supersymmetry and supergravity give a natural unification of matter and force. Wesson (1983, 84) and Reddy (1999) have studied several aspects of five dimensional space-time in variable mass theroy and bimetric theory of relativity respectively. \\
Due to the virial theorem we have the following connection between kinetic and potential energies of an isolated gravitating system
\beq										
2T + V = 0
\eeq
The potential energy of a static celestial body can be given using Newton's law of gravitation by,
\beq
 V \sim G\frac {M^2}{R_s}
\eeq					
while the kinetic energy can be expressed using the pressure which balances configuration by,
\beq										
T \sim  p R^3_s						
\eeq
Here, we denote by $R_s $ the radius of the configuration. So we find a qualitative expression widely used in astrophysics
\beq
M^2 \sim \frac{1}{G}pR^4_s  				
\eeq
Avakian (1990) discussed the exact expression of the connection between the mass, pressure and volume for static spherically symmetric configuration in the framework of four dimensinal space-time. In this paper we derived the mass square of a celestial body which is  represented as an integral over the pressure distributions taken over the volume of the body in the frameworks of five dimesional spac-time.

\subsection{The mass formula for static configurations in five dimesional space-time}

The five dimensional line element corresponding to a static spherically symmetric configuration in isotropic coordinates is given as
\beq
ds^2  = c^2 e^\nu dt^2 - e^\lambda [ dr^2 + r^2 ( d\theta_1^2 + sin^2 \theta_1 d\theta_2^2 + sin^2\theta_1 sin^2\theta_2 d\theta_3^2 )]	
\eeq
where $ \nu $ and  $ \lambda $ are functions of $ r $-only.
The five velocity has vanishing spatial components $ u^1 = u^2 = u^3 = u^4  = 0 $ and  $ u^0 u_0 =1  $ so we have the following nonvanishing components of the energy momentum tensor,
\beq
  T^0_0 = \rho, \: \: \; \; T^1_1 = T^2_2  =  T^3_3 = T^4_4 = - p
\eeq
where $ p $ denotes the pressure and  $ r  $  {\large }is the matter density of the configuration.
The Einstein field equations in five dimensions are
\beq
 R^j_i =  -\frac{8 \pi G }{c^4} [ T^j_i - \frac{1}{3} \;\; T ] 
\eeq
From (5) and (6), equation (7) can be expressed as,
\beq
R^0_0 = -\frac{8 \pi G}{c^4} [ \frac{2}{3} \rho c^2 + \frac{4}{3}p]
\eeq
\beq
R^1_1 = R^2_2 = R^3_3 = R^4_4 =  -\frac{8 \pi G}{c^4} [- \frac{1}{3} \rho c^2 + \frac{1}{3}p]
\eeq
where   				
\beq
 R^0_0 =  -\frac{1}{2} e^{-\nu}[ \nu''+\lambda'\nu' +\frac{1}{2} \nu'^2 + 3 \frac{\nu'}{r}]
\eeq
\beq
R^1_1 = -\frac{1}{2} e^{-\lambda}[ 3\lambda''+3 \frac{\lambda'}{r} -\frac{1}{2} \nu' \lambda'  +\nu'' + 3 \frac{1}{2}\nu']
\eeq
\beq
 R^2_2 = R^3_3 = R^4_4 = -\frac{1}{2} e^{-\lambda}[ \lambda''+5 \frac{\lambda'}{r} + \lambda'^2 + \frac{1}{2} \nu' \lambda'  +\frac{\nu}{r}]
\eeq
and prime denotes derivative with respect to $ r.$
 Again from (8) and (9)
\beq
R^0_0 + 2R^4_4 = - \frac{16 \pi G p}{c^4}
\eeq
     
From (10) and (12), equation (13) can be written as
\beq
\frac{16\pi Gp}{c^4} e^\lambda = \lambda'' +\frac{\nu''}{2} +\lambda'^2  + \frac{\nu'^2}{4} + \nu'\lambda' +5 \frac{\nu'}{r}
\eeq
Multiply both sides by $ r^3 e^{(\lambda+\frac{\nu}{2})}$
\beq
\frac{16\pi Gp}{c^4}  r^3 e^{(2 \lambda+\frac{\nu}{2}) } = [  r^3 e^{(\lambda+\frac{\nu}{2})} (\lambda' +\frac{\nu'}{2} + \frac{1}{r})]^{'}	- 2e^{(\lambda + \frac{\nu}{2})}
\eeq
Multiply both sides by $ r^2$
\beq
\frac{16\pi Gp}{c^4}  r^5 e^{(2 \lambda+\frac{\nu}{2}) } = [  r^5 e^{(\lambda+\frac{\nu}{2})} (\lambda' +\frac{\nu'}{2}]'
\eeq
After integrating (16) for  $ r $ from zero to infinity taking into account that $ p(r) = 0 $  for $ r ³ \ge r_s, $ where $r_s  $is the radius of configuration we get,
\beq
\frac{16\pi Gp}{c^4} \int_0^{r_s} p\;r^5  e^{(2 \lambda+\frac{\nu}{2}) } dr =  r^5 e^{(\lambda+\frac{\nu}{2})} (\lambda' +\frac{\nu'}{2})
\eeq

The five dimensional exterior solution of Einsteins equations in isotropic coordinates are given by,
\beq
e^\lambda = ( 1 + \frac{r_g}{4r^2})^2 , \;\:\:\:\:\:  e^\nu =  \frac{(1-  \frac{r_g}{4r^2})^2}{(1+  \frac{r_g}{4r^2})^2}
\eeq
 			 		
where $r_g = \frac{2GM}{c^2} $ is the gravitational radius of the body and  M denotes mass of the body.

Using (18), equation (17) reduces to
\beq
\frac{16\pi Gp}{c^4} \int_0^{r_s} p\;r^5  e^{(2 \lambda+\frac{\nu}{2}) } dr = \frac{r_g^2}{4}
\eeq
using $r_g = \frac{2GM}{c^2}$, equation (19) is expressed as 				
\beq
M^2 = \frac{16\pi}{G} \int_0^{r_s} p\;r^5  e^{(2 \lambda+\frac{\nu}{2}) } dr
\eeq
\subsection{The mass formula for static configurations in Kaluza-Klein space-time}

Consider the line element in the form,
\beq
ds^2 = c^2 e^\nu dt^2 - e^\lambda [ dr^2 + r^2d\theta^2 + r^2sin^2\theta d\phi^2 ] - e^\mu dy^2
\eeq	
where $ \nu \; \& \lambda $  are functions of $r $ only and $y$  is a  Kaluza-Klein parameter.

The nonvanishing Ricci tensors for the metric (21) are
\beq
R^0_0 = -e^{-\lambda} [ \frac{\nu''}{2} +\frac{\nu'^2}{2} +  \frac{\lambda' \nu'}{4}+  \frac{\nu' \mu'}{4}+  \frac{ \nu'}{4}]
\eeq
 \beq
R^1_1 = -e^{-\lambda} [ \frac{\nu''}{2} + {\lambda}''+ \frac{\mu''}{2} -  \frac{\lambda' \nu'}{4}-  \frac{\lambda'  \mu'}{4}+  \frac{ \lambda'}{r} + \frac{\nu'^2}{4} +\frac{\mu'^2}{4}]
\eeq
\beq
R^2_2 = R^3_3 = -e^{-\lambda} [ \frac{\lambda''}{2} + \frac{3\lambda'}{2r} +  \frac{ \nu'}{2r}+  \frac{\mu'}{2r}+  \frac{ \lambda' 
\nu'}{4} + \frac{\lambda'\mu'}{4} +\frac{\lambda'^2}{4}]
\eeq
\beq
R^4_4 = -e^{-\lambda} [ \frac{\mu''}{2} +\frac{\mu'^2}{4} +  \frac{\mu' \nu'}{4}+  \frac{\mu' \lambda'}{4}+  \frac{ \mu'}{4}]
\eeq 	
From equations (6) and (7), we have
\beq
 R^0_0 + 2R^2_2 = -\frac{16 \pi G}{c^4} p
\eeq						

using (22), (24) and (25), equation (26) can be expressed as,
\beq
 \frac{1}{2}[{\nu''} + {\lambda''} +{\mu''}]+ \frac{1}{4}[{\nu'^2} +  {\lambda'^2} + {\mu'^2} +{\lambda'  \nu'}+ { \lambda' \mu'} + +{\mu' \nu'}]  + \frac{3}{2r}[{\nu'} + {\mu'}+{\lambda'}] = - \frac{16 \pi G}{c^4}  \; p \; e^\lambda
\eeq
{ \bf For a particular case $\mu=0 :$}\\

Equation (27) reduces to,
 \beq
 \frac{1}{2}[{\nu''} + {\lambda''} ]+ \frac{1}{4}[{\nu'^2} +  {\lambda'^2}  +{\lambda'  \nu'}]  + \frac{3}{2r}[{\nu'} +{\lambda'}] = - \frac{16 \pi G}{c^4} \; p \; e^\lambda 
\eeq
 
multiply both sides by  $ r^2 e^{(\lambda + \frac{\nu}{2})} $
\beq
 \frac{16 \pi G} {c^4} p r^2  e^{(3 \lambda + \frac{\nu}{2})} =  [ r^2  e^{(3 \lambda + \frac{\nu}{2})}(\frac{\nu' +\lambda'}{2} +\frac{1}{r}) ]' - e^{(\frac{\lambda +\nu}{2})}
\eeq                                    
multiply both sieds by $r$ 
\beq
 \frac{16 \pi G} {c^4} p r^2  e^{(\frac{3 \lambda + \nu}{2})} =  [ r^3 ( e^{ (\frac{\lambda + \nu}{2})})']'
\eeq
Integrating from $ r = 0 $  to infinity taking into account that $ p(r) = 0 $  for $ r \ge r_s, $ where $r^s $ is the radius of configuration,
\beq
 \frac{16 \pi G} {c^4} \int^{r_s}_0 p\; r^3 e^{(\frac{3 \lambda + \nu}{2})} dr  =  r^3 ( e^{(\frac{\lambda + \nu}{2})})'
\eeq
 				
From the five dimensional Schwarzschilds solution in isotropic coordinates (Wessons (1999)) .

\beq
e^\lambda = ( 1 + \frac{r_g}{4r})^4 , \;\:\:\:\:\:  e^\nu =  \frac{(1-  \frac{r_g}{4r^2})^2}{(1+  \frac{r_g}{4r^2})^2}
\eeq		 				
where $r_g = \frac{2GM}{c^2} $ is the gravitational radious of the body.

Using (32), equation (31) reduces to 

\beq
 \frac{16 \pi G} {c^4} \int^{r_s}_0 p\; r^3 e^{(\frac{3 \lambda + \nu}{2})} dr  =  \frac{r_g^2}{8}
\eeq                                 

using   $r_g = \frac{2 G M}{c^2}, $ equation (1.33) can be expressed as
\beq
 M^2 = \frac{32 \pi } {G} \int^{r_s}_0 p\; r^3 e^{(\frac{3 \lambda + \nu}{2})} dr 
\eeq                                 
The above expressions of the mass  in Kaluza-Klein theory is similar to the expression of the mass obtained earlier by Beylar-et-al (1995) in four dimensional space-time.
\subsection{Conclusion:}
In the present work, we have derived the new formulae  for the mass of spherically symmetric stellar configurations in the frameworks of five dimensional space-time. We observed that the mass square of  static spherically symmetric celestial body is connected to the pressure distribution in the framework of five dimesional space-time. It is also observbed that in Kaluza-Klein space time  the expressions of the mass are similar as that of expression of the mass obtained by Beylar-et-al (1995) in four dimension. This new mass  formulae is of importance especially in numerical calculations of  the masses of celestial bodies.  .

\bibliographystyle{plain}

\small
\vskip 1pc
{\obeylines
\noindent G S Khadekar
\noindent Department of Mathematics 
\noindent Nagpur University, Mahatma Jyotiba Phule Educational Campus 
\noindent Amravati Road, Nagpur- 440 010 (MS) (INDIA)
\noindent e-mail: gkhadekar@rediffmail.com
\vskip 1pc
{\obeylines
\noindent K S Adhao
\noindent Department of Mathematics 
\noindent Amravati University, Amravati (MS) INDIA
\noindent e-mail: ati-ksadhav@sauharnet.in
\vskip 1pc
\noindent H R Ghate
\noindent Department of Mathematics 
\noindent Jijamata Mahavidhyalaya 
\noindent Buldhanar (INDIA)
\noindent Submitted: October 18, 2002
\end{document}